\documentclass[prd,twocolumn,letterpaper]{revtex4}
\usepackage{graphicx} 

\begin{document}

\title{Mechanical and Acoustic Studies of \\
Deep Ocean Glass Sphere Implosions}
\author{P.~W.~Gorham, M.~ Rosen, J.~W.~Bolesta, J.~G.~Learned,
J. Reise}
\affiliation{Department of Physics and Astronomy, University of Hawaii \\
2505 Correa Rd., Honolulu, HI, 96822}


\begin{abstract}
Three deep-ocean implosions of 43 cm diameter 
glass instrument housings, made by Benthos Inc., 
were studied to determine the effect on their associated
cabling and moorings. 
High resolution acoustic profiles were also measured for
two of the three implosions, allowing us to infer some of the
dynamics and kinematics of the events. The mechanical forces 
on the ancillary mooring hardware during the
entire implosive/explosive event were found to be most probably
dominated by the explosive shock wave following the initial infall.
A syntactic float at a distance of 16 m from the implosion center
was probably shattered by such a shock wave, but 3 glass instrument
housings apparently survived within a distance of 6 m from the same
implosion.
\end{abstract}
\maketitle

\section{Introduction}

Although many thousands of large spherical glass instrument housings 
and floats have been 
deployed in the ocean to depths of 6 km or more, the statistics on
failure of these housings under pressure are not well established. 
Sphere failures during test cycles after manufacture are very rare,
and are most often traceable to mishandling rather than inherent flaws
in the glass itself. Records of sphere failure {\em in situ} are
usually very incomplete; such failures often lead to loss of the instrument
package and therefore no knowledge of the cause of the loss.

Because the failure rate in documented cases has been known to be
primarily due to handling problems,
it is virtually impossible to accurately 
estimate the likelihood of sphere failure
in a large array of many hundreds of instrument spheres deployed in
the deep ocean over many years' time. Thus knowledge of the effects
of such an implosion is useful in determining what measures must
be taken to ensure survival of other instruments in the vicinity.
(A review of 
the use of spherical glass housings in the deep ocean may be found in
Raymond (1975)). 

At present a number of experiments are in the process or planning stages
for deployment of large arrays of hundreds of galss sphere instrument
housings in the deep ocean for the study of high energy particle interactions
which can yield information about neutrinos, one of the most elusive of
all subatomic particles. Such arrays will use various techniques for 
suspending the instrument housings via tensile riser cables or structures,
which must also contain telemetry cables.
Because of the difficulty in assessing the likelihood of sphere
failure, and the knowledge that at the working array depths each sphere
has a potential energy of $\sim 1.5 \times  10^6$ Joule, we devised
an experiment which could test the viability of the tensile
mooring and the electro-optical power and telemetry riser
cables for survival in the
event of implosion. Since destruction of any of these three components could
lead to failure of a larger portion of the array, the 
experiment was designed to test a number of possible configurations
which might mitigate such damage.

In addition to the photo-optical sensors in the array which are designed
to detect Cherenkov emission from energetic charged particles which may
be the interaction products of a primary neutrino, the arrays envisioned
will also contain an acoustic sub-array which is designed to track the
positions of all of the sensors to $\sim 1$ cm precision in real time.
Because of the potentially 
damaging strength that an acoustic pulse from an implosion might have
for the hydrophone array, we also made surface recordings of the
sound intensity. Although such measurements have been made in the past
for deep-ocean implosions of similar housings (Orr and Schoenberg 1976),
we have improved on the sampling interval by more than a factor of four;
thus such measurements are of interest to those studying ocean acoustics,
as well as the dynamics of implosive events.

\section{Experimental Setup}
  
As was shown in Orr and Schoenberg (1976), based on work done with 
the Benthos Corporation, glass instrument housings may be
prepared in such a way that they will implode at a predetermined depth,
by grinding a flat on a section of the sphere exterior. Three such spheres
were used for this experiment, with the flat sizes chosen to implode
at a depth of $\sim 4000$ m. A cross-section of such a sphere
is shown in Figure 1, with the wall thickness at
the flat center as the indicated dimension. 
The ratio of this to the sphere radius is denoted
the flat-thickness ratio (FTR).
  
\begin{figure}
\centerline{\includegraphics[width=3.35in]{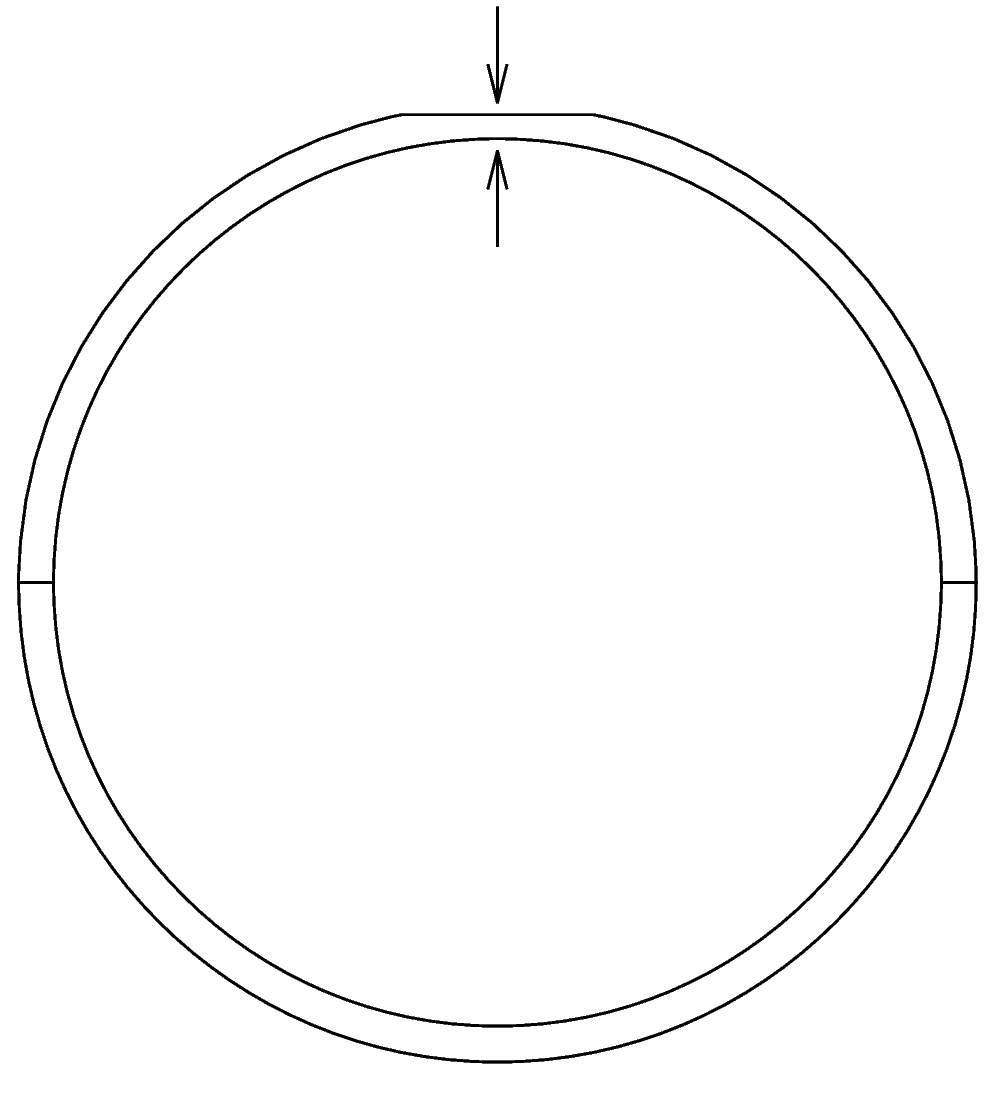}}
\caption{Implosion sphere geometry. The indicated dimension is the
wall thickness at the flat; the ratio of this to the sphere radius
determines the flat-thickness ratio (FTR).}
\end{figure}

It is standard practice in installation of oceanographic moorings to
encase the instrument spheres within plastic hardhats. These are
usually made of polyethylene or ABS plastic.
The requirement for transparency in photosensor arrays has led us to the
use of acrylic hardhats, primarily because of the superior ultraviolet transmissivity of pure acrylic compared to most other clear plastics.
However, acrylic
hardhats are somewhat less mechanically robust than some other choices.
For this experiment, a set of acrylic hardhats used in a previous
ship-tethered experiment were used (O'Connor 1990).

The initial experimental setup is shown in Fig. 2 (a). An aluminum
frame made of T-bar stock was used to provide a tensioning jig so that
the module vehicle could simulate the expected tension in the
actual array ( $\sim 5$ KN per side). 
The side view of the implosion module does not
show the mooring geometry, but this consists of a pair of kevlar risers,
$\sim 12$ mm diameter, with fiber optic and electrical cable bundles on both
sides of this dual riser, each with diameters of order 2-3 cm. For the
initial experiment, the kevlar risers and electro-optical cable bundle
were encapsulated within a channel in the acrylic hardhat with some 
clamping pressure applied by the acrylic. 
Cross members in the tensioning frame were used to rotate the module
$90^{\circ}$ relative to the plane of the tensioning frame. 
This was done to minimize the effects of the T-bar frame on the
implosion dynamics near the dual risers.

\begin{figure*}
\centerline{\includegraphics[width=6in]{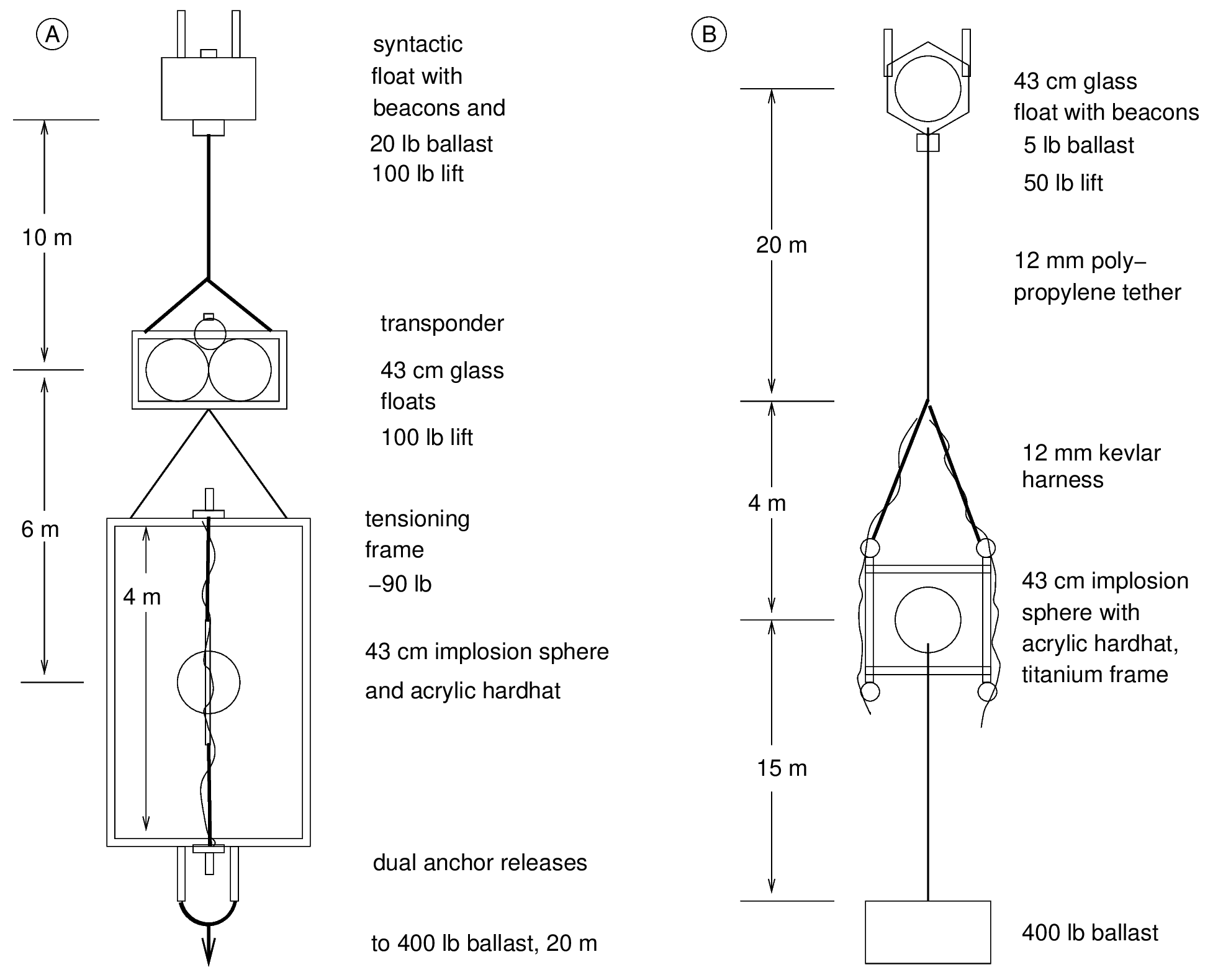}}
\caption{(A) A schematic of the setup for the first implosion test.
The wavy lines indicate schematically the mock electro-optical riser
elements, which were not under tension in the test.
(B) Schematic for the second test. }
\end{figure*}

The experiment was designed as a free-fall vehicle, with
two parallel mechanical anchor releases that could be
activated either by loss of a small tether attached to the acrylic
hardhat, or by a corrodible link. A concrete ballast with 
400 lb wet weight was used to overcome the dual bouyancy packages:
one using a pair of 43 cm glass floats (with a smaller transponder 
sphere mounted above them), the second a large syntactic foam
float with approximately 20 lb lead counterballast to provide a righting
moment.

As we will discuss in the following section, this initial package
was lost in spite of the redundancy of releases and buoyancy
in its design. A second package was assembled from the spares of the
first package, as shown in Fig. 2 (b). This package did not include
a tensioning frame, but did employ a protective titanium "picture frame"
around the acrylic hardhat. This frame was a design chosen to provide
as much protection as possible for the electro-optical 
riser cables. This was accomplished
by terminating the kevlar tensile members at the corners of the Ti frame,
while leading the electro-optical cables past the module on the 
outside face of a T-section of 1/8" titanium with the stem of the
T facing the module. The twin halves of the acrylic hardhat then
sandwiched the T-stock at this stem section, and were through-bolted
to it with 1/4" bolts. Some of the bolts used for this were grade-8
high-tensile strength steel; others were mild steel. The titanium
T-sections were stitch-welded 6Al4V alloy, with a tensile strength
of order 120-130 KPSI.

The acoustic recording system consisted of a Marantz professional Digital
Audio Tape (DAT) recorder, sampling at 48 KHz on dual channels with
individually settable gains. The channels were set to attenuations of
-20 dB and -60 dB relative to the nominal hydrophone gain.
The hydrophone used was an Ocean and Atmospheric Science
E-series model, with a sensitivity of
-87 dB re 1 V per $\mu$Pa. The variation of this sensitivity with frequency
will be discussed below.
The hydrophone was deployed on a weighted line with approximately
20 m of cable, at a depth of $\sim 5$ m below the surface. Because of the
sea state (Beaufort II-III) 
during most of the operations, the measurements were made with a 
slight forward weigh on to keep the vessel 
(a 43 foot fishing boat) head to the seas.

\section{Observations}
  
The tests were conducted approximately 20~km from the west coast of
Hawaii Island, 
within a radius of 1.5 miles of $19^{\circ} ~39'$ N $156^{\circ} ~20'$ W,
during the period 26-28 June 1992, in water depths ranging from 4.5-4.7 km.
Three different implosions were successfully achieved at depth: the first
with the package shown in Fig. 1(A), the second with that shown in
Fig. 1(B), and the third an expendable package consisting only of a glass ball
within a thin rubberized cloth bag tethered to a 400 lb wet-weight
ballast.

\subsection{First test}
  
The initial package carried a Benthos XT-6000 
transponder by which slant ranges to the
descending instrument could be determined. After the package was released,
the descent rate was determined to be $\sim 1.4 ~\rm{m} ~\rm{s}^{-1}$, and this
rate continued until the implosion occured after a span of
$\sim 52$ min. Shortly prior to the implosion, the hydrophone cable became
entangled in the vessel`s propellor, and it was not recovered and redeployed
in time to record the first acoustic profile. However, divers present in
the water at the time clearly heard the signal, and a transponder range
was obtained immediately. This, combined with satellite navigation
(GPS) positions obtained at the time, gave an implosion depth of
$4140 \pm 70$ m, using a standard curve for the
velocity of sound obtained in the same area (O'Connor 1990).

After the implosion occurred, transponder ranges were taken to determine
whether the ballast releases had functioned successfully, and it
was found that the package continued its descent. The descent rate had,
however, slowed to $\sim 0.5 ~\rm{m} ~\rm{s}^{-1}$. This indicated that some
significant change in net buoyancy had occurred, not due to the loss of
buoyancy of the implosion sphere, since this would only increase the 
descent rate if the main ballast remained present. The package continued 
its descent until it rested on the bottom at a depth of approximately
4640 m. The survival of the transponder, contained within a smaller glass 
sphere directly adjacent to the 43 cm glass floats, strongly indicates that
no sympathetic implosion took place among these floats, at a distance
of $\sim 6$ m from the center of the imploding sphere.

During a return to station to compensate for drift approximately 5 hours
following the implosion, one half of the top syntactic float was 
encountered on the surface. This was recovered, and examination of it
indicated that it had been shattered, and the recovered section did not 
carry the righting ballast or the stainless steel attachment shaft.
Two other smaller pieces of the float were also seen and recovered, both
with similar shattered appearance. We surmise that the shatter of the float
occurred as a result of the implosion-induced shock wave impact on the
float, which had a flat surface area of order $1/3 ~\rm{m}^2$ facing
directly at the imploding sphere, although it was $\sim 16$ m distant
from it.

Although the remaining buoyancy of the two glass floats was 
estimated to be just sufficient to return the package to the surface, the
fact that the syntactic float's righting ballast and associated hardware
remained attached to
these spheres, while the syntactic float was entirely lost, probably accounts
for the failure of this package to return. The change in fall rate mentioned
previously suggests strongly that the anchor did release. Given the
results of the following test it is difficult to imagine how the triggering
scheme, which was based on fracture of the acrylic hardhat, could have
failed to operate successfully.

\subsection{Second Test}

Once it became apparent that the secondary corrosion releases of the
system were also unlikely to aid in returning the first package, we
assembled a second free-fall 
package from the remaining spares, intending to still
gather information on the implosion damage if possible. This package
could not simulate the tensile riser tension, but did provide a reasonable 
simulation of the electro-optical riser configuration, since this is
not under tension in the array design.
As mentioned above, the acrylic hardhat in this case was 
captured within a rigid ``picture
frame'' of titanium which was designed to survive even if the
hardhat were destroyed. The release consisted of a single-point attachment of
the ballast to the acrylic hardhat, with a shock cord in parallel to 
minimize the shock loading on the acrylic.

This package was successfully deployed on our second attempt, after one
anchor pre-trip. The implosion was recorded
on the DAT system after an elapsed interval of 33 min, with an
estimated depth of $4280 \pm 50$ m, based on later analysis of
the acoustic echos and the known bottom depth. The return package
was sighted and recovered within another hour. It was found that the
acrylic hardhat and glass sphere were completely destroyed, with only two 
small shards of acrylic, approximately 1/2 cm in diameter, found
still clamped behind two of the bolts. Only one of the titanium
frame crossmembers remained, and this one 
had been detached on one side. An 
examination of the broken welds suggested that they were broken in tension
rather than compression.

All of the steel through-bolts that held the acrylic to the Ti frame
were found to be bent, some as much as $\sim 10^{\circ}$,
by shear loading on each side 
of the titanium T stem. Later examination of the rims of the titanium frame
holes that had contained the high-tensile steel bolts
showed conclusively that the bending of these bolts was due to forces directed
radially outward from the center of the frame, indicating the 
the dominant forces were explosive rather than implosive. The titanium frame
had yielded  typically $\sim 1$ mm in these holes, but the vertical titanium 
section had retained its integrity as a tensile member of the mooring.

Visual examination of the kevlar and electro-optical riser cable bundles
showed no damage, beyond a small nick in one of the kevlar splices which
was attributed to its proximity to the sharp surface of the broken
crossmember weld. No shards of glass or acrylic were evident on any of these
riser surfaces, and none were found subsequently within
the kevlar yarns or cable bundles. 
High density polyethylene, and polyvinyl chloride thimbles
used to terminate the kevlar 
also showed no evident damage under visual inspection.

\subsection{Third test}
  
A third implosion test was conducted without any hardhat to determine a
baseline acoustic profile for normalizing the profile obtained with the
hardhat and frame. This was conducted successfully, with an elapsed
descent time of 29 min. The implosion depth was later estimated to be
$4470 \pm 50$ m, by bottom echos. Thus the final implosion was the 
deepest of the three, with almost 10\% higher potential energy available than
the shallowest implosion.

\section{Results}

\subsection{Implosion Depth Relation}
  
Orr and Schoenberg (1976) found an empirical relation between the FTR
of a ground sphere and the implosion depth. Based on
laboratory experiments with a pressure vessel, the dependence of the
implosion depth on the FTR was found to be nearly linear. However,
the fit of this laboratory curve to the actual field data
of Orr and Schoenberg was imperfect, suggesting that the curve might be
different under deep sea conditions as compared to a pressure vessel.
Because of this possibility, the spheres for our test were ground 
to FTRs extrapolated from only the field data of Orr and Schoenberg.
The FTR required to fit the laboratory curve for implosion
depths of 4 km could actually lead to real implosions as deep as
4.8 km, beyond our site depth, if the deep-ocean behavior is different 
from that in the laboratory.

\begin{figure}
\centerline{\includegraphics[width=3.5in]{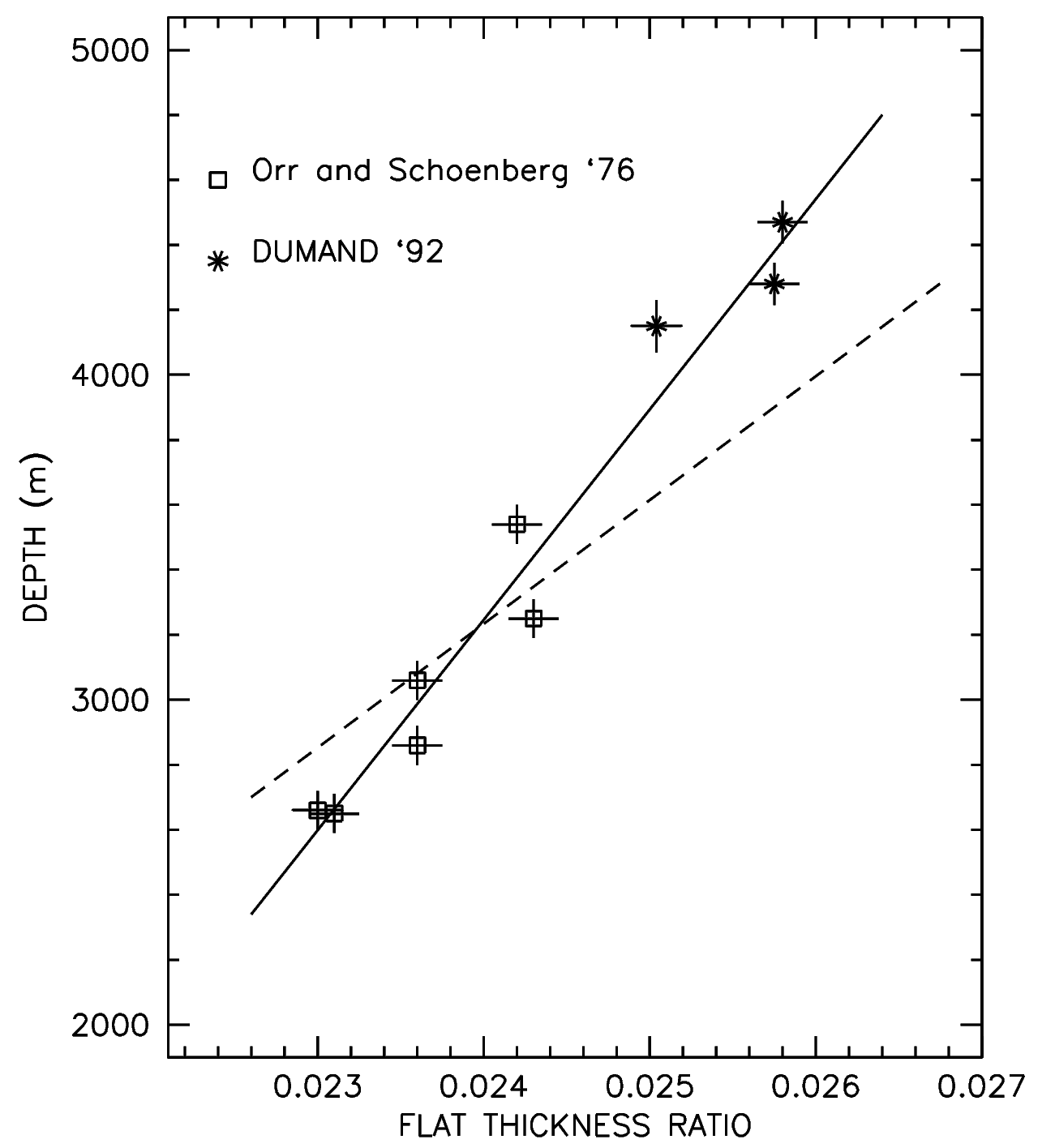}}
\caption{Implosion depth vs. the Flat Thickness Ratio (FTR) for
nine 43 cm sphere deep-ocean implosions. The solid line is a fit to the deep
ocean data; the dashed line a fit to Benthos pressure-vessel data for
comparison.}
\end{figure}

Figure 3 shows our results and those of Orr and Schoenberg (1976), along
with two curves: the dashed line indicating the best fit for
the laboratory data of Orr and Schoenberg, and the solid line
a fit to only the deep ocean data. Our additional data clearly
indicate that the laboratory tests do not adequately describe the
dependence for the deep ocean implosions. However, the data do
fit well to a straight line and it appears that glass spheres may be
prepared for implosion to a precision of order 50-100 m depth for such
tests.

The empirical dependence of the implosion depth on FTR 
for 43 cm housings similar to the Benthos Inc. type is found to be:
\begin{equation}
D ~(\rm{meters})  =  7.46 \times 10^4 ~(FTR - 0.0230) ~+~ 2600 
\label{ftr-eqn}
\end{equation}
valid over the range 2 to 5 km and an FTR range of 0.0230 to 0.0265.
The difference between this result and that measured in the lab may
be attributable to many factors; one obvious one is the temperature
difference, which could increase the relative strength of the glass 
in the ocean. A second factor which is probably more significant is
the smaller pressure reservoir available in a laboratory pressure
tank as compared to the deep ocean.

\subsection{Acoustic Profiles}
  
The acoustic profiles are shown in Figure 4, with the solid line
indicating the profile of the deeper implosion, which
had no hardhat or frame; and the dashed line
that of the shallower implosion, which had both. 
As mentioned above, the acoustic profile of the first test was not
recorded, but both the second and third tests were successfully
measured. The high-gain channel in both cases was saturated by the peak of
the shock wave; thus Fig. 4 shows only the profiles recorded in the low-gain
channel. The shallow depth of the hydrophone gave a prompt surface
reflection which confuses the profiles $\sim 4$ ms after the implosion onset;
these reflections are not included in Figure 4.
  
\begin{figure}
\centerline{\includegraphics[width=3.4in]{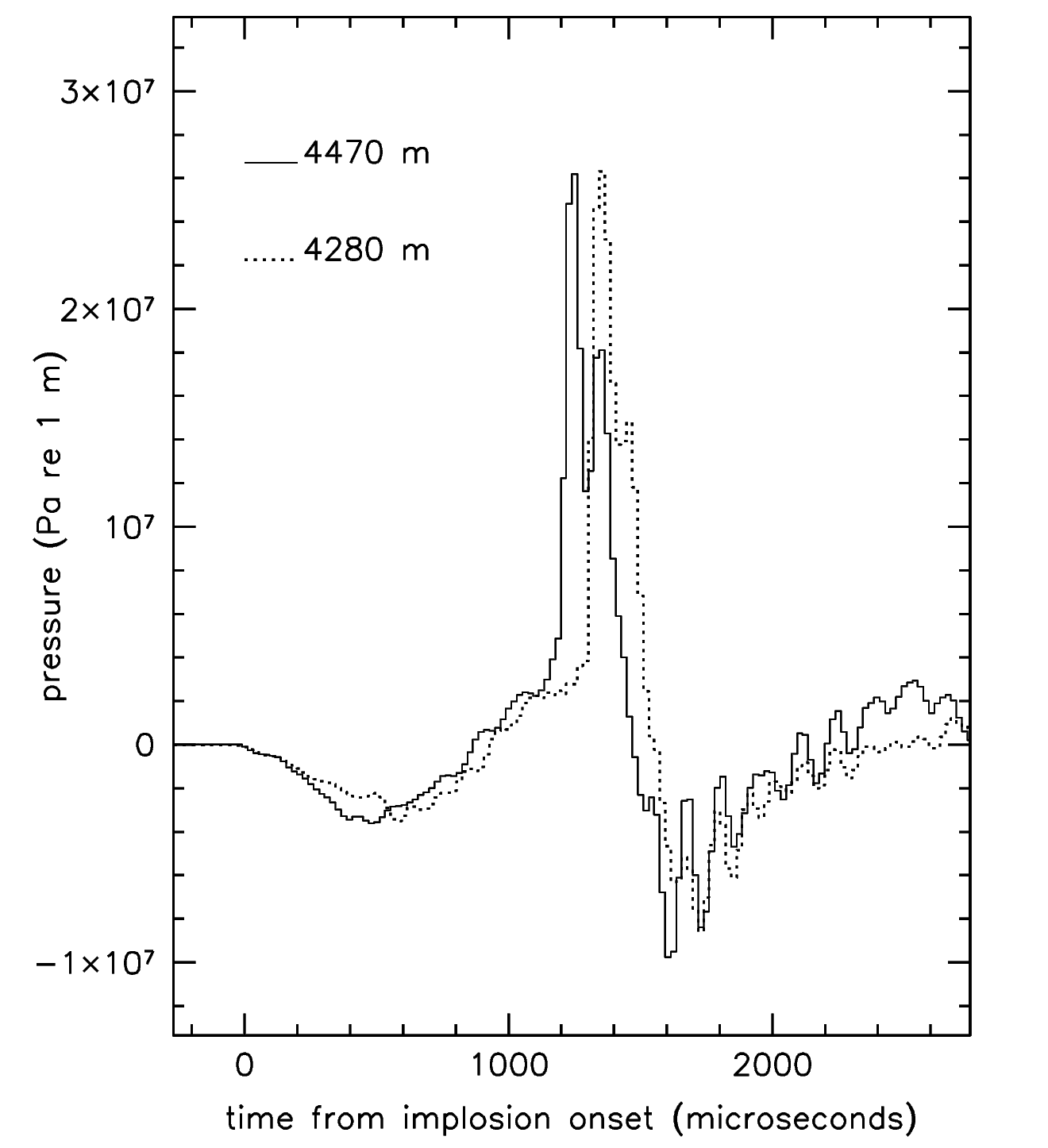}}
\caption{Time profiles of the two implosions for which they were
recorded. The apparent oscillation after the positive peak is due
to a hydrophone artifact. The pressure has been referenced to 
ambient pressure at a distance of 1 m from implosion center.}
\end{figure}

The calibration of the pressure scale is shown in Pascals referenced to
1 m radius from the source, assuming spherical spreading. The
absolute scale was determined from the results of
Orr and Schoenberg (1976), 
using a fit to the magnitude of the underpressure pulses for the six 
43 cm spheres they tested. The leading underpressure pulse
is unaffected by acoustic attenuation since it is low frequency;
it was found that the magnitude of this pulse was well-fit by
a linear relationship with depth, 
allowing us to extrapolate to the implosion depths of our data.

Unfortunately, the hydrophone used to record the acoustic profiles was
later found to have a local resonance over the frequency range
6-10 kHz, and reduced sensitivity over the range 6-12 kHz.
We have not yet attempted to measure the complex transfer function of
the hydrophone, which might allow us to fully correct for these effects.
However, such transfer functions can also have amplitude dependence which
makes it difficult to calibrate highly impulsive events. For the present
it is best to use the data in the range of 6-12 kHz with caution; however,
the hydrophone manufacturer claims that the sensitivity is uniform over
other portions of the spectrum up to $\sim 25$ kHz.

The acoustic energy may be estimated by
(Urick 1967): 
\begin{equation}
E   =   {{ 4 \pi } \over{ \rho c}} \int_0^T P^2 (t) ~dt   
\label{energy-eqn}
\end{equation}
where $P (t)$ is the measured pressure (referenced to $r=1$ m here), 
$\rho c$ the density-velocity product for seawater, the $4 \pi$
factor accounts for solid angle, and T is the duration of the acoustic event. 
Figure 5 shows the acoustic energy fluence
as a function of time, and Figure 6 the cumulative total acoustic energy, both
estimated from the surface measurements, 
assuming spherical symmetry.

\begin{figure}
\centerline{\includegraphics[width=3.2in]{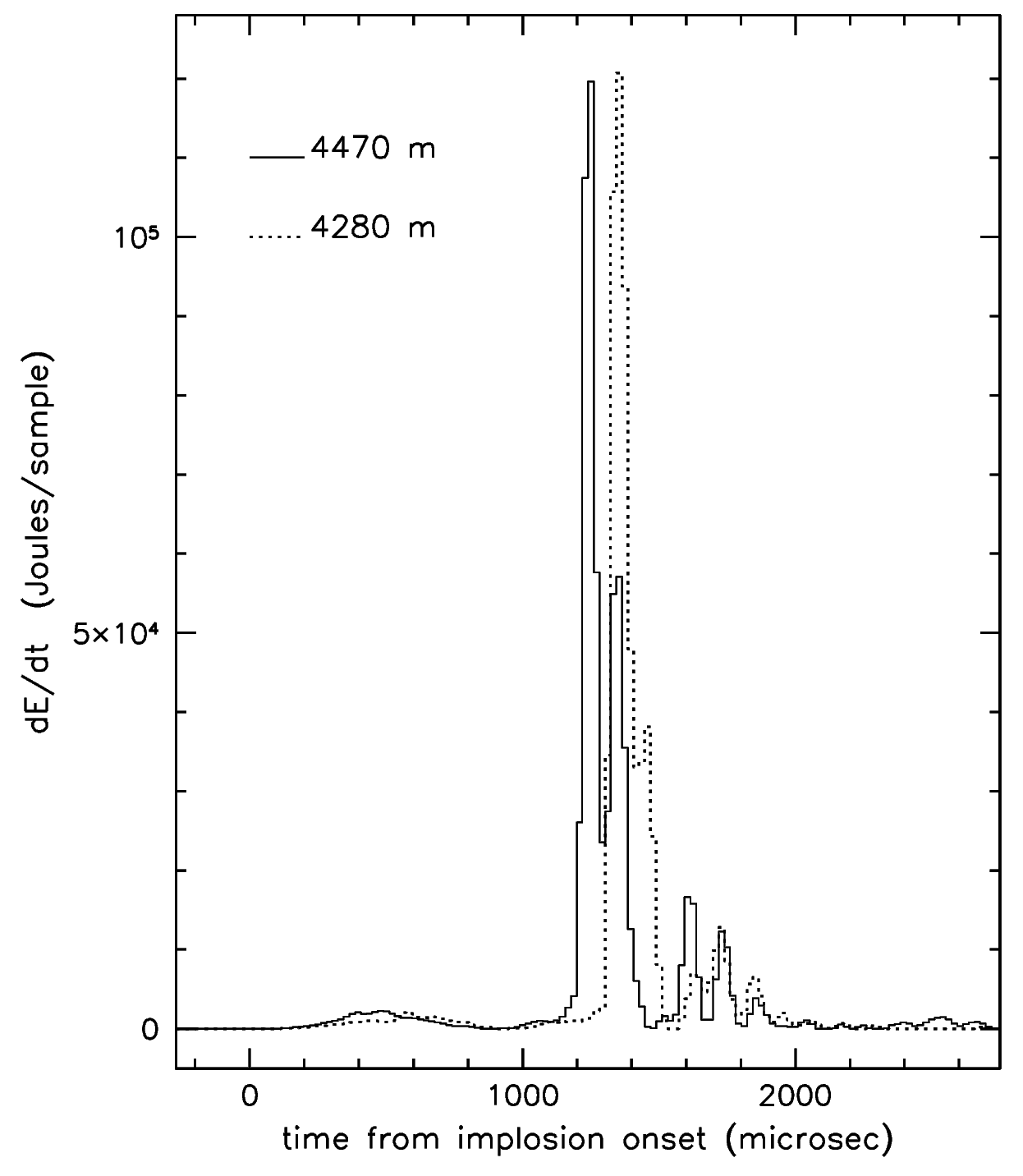}}
\caption{Total acoustic energy fluence of the two events,
assuming spherical symmetry. The sample duration
was about 20 microseconds.}
\end{figure}

\begin{figure}
\centerline{\includegraphics[width=3.35in]{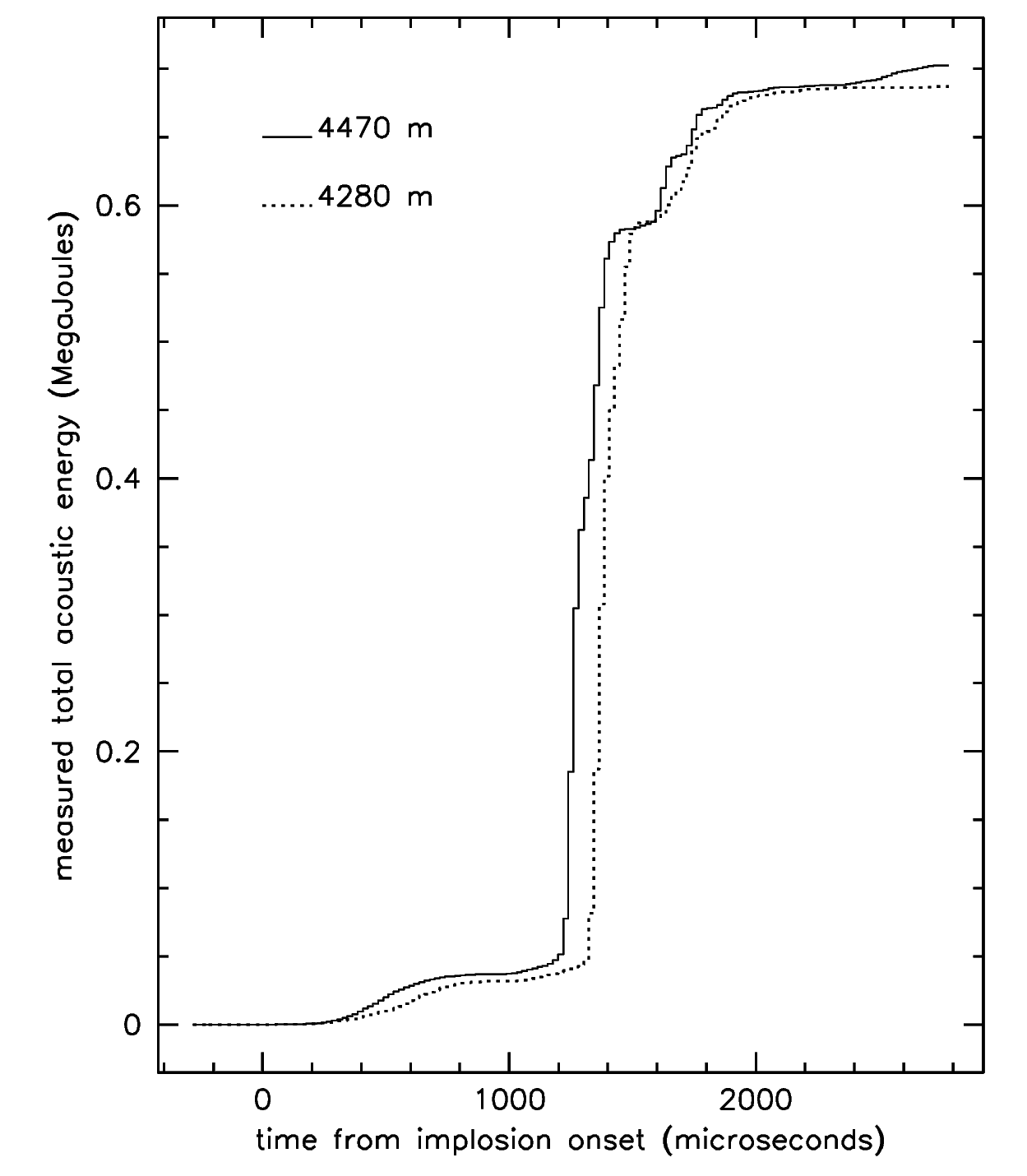}}
\caption{Cumulative total acoustic energy for the events, also assuming spherical
symmetry.
}
\end{figure}

The acoustic power spectra of the two pulses are shown in Figure 7.
A number of hydrophone artifacts are present as discrete resonances
apparent both in the time series and power spectra at $\sim 7$ and $\sim 9$
kHz, and possibly at their first overtones as well. The spectrum has not been 
corrected for this hydrophone response function. However, we have
corrected for the known water-column attenuation, which 
is approximately quadratic in frequency above $\sim 5$ kHz in our data, and
amounts to $\sim 10$ dB at the Nyquist limit.

\begin{figure}

\centerline{\includegraphics[width=3.35in]{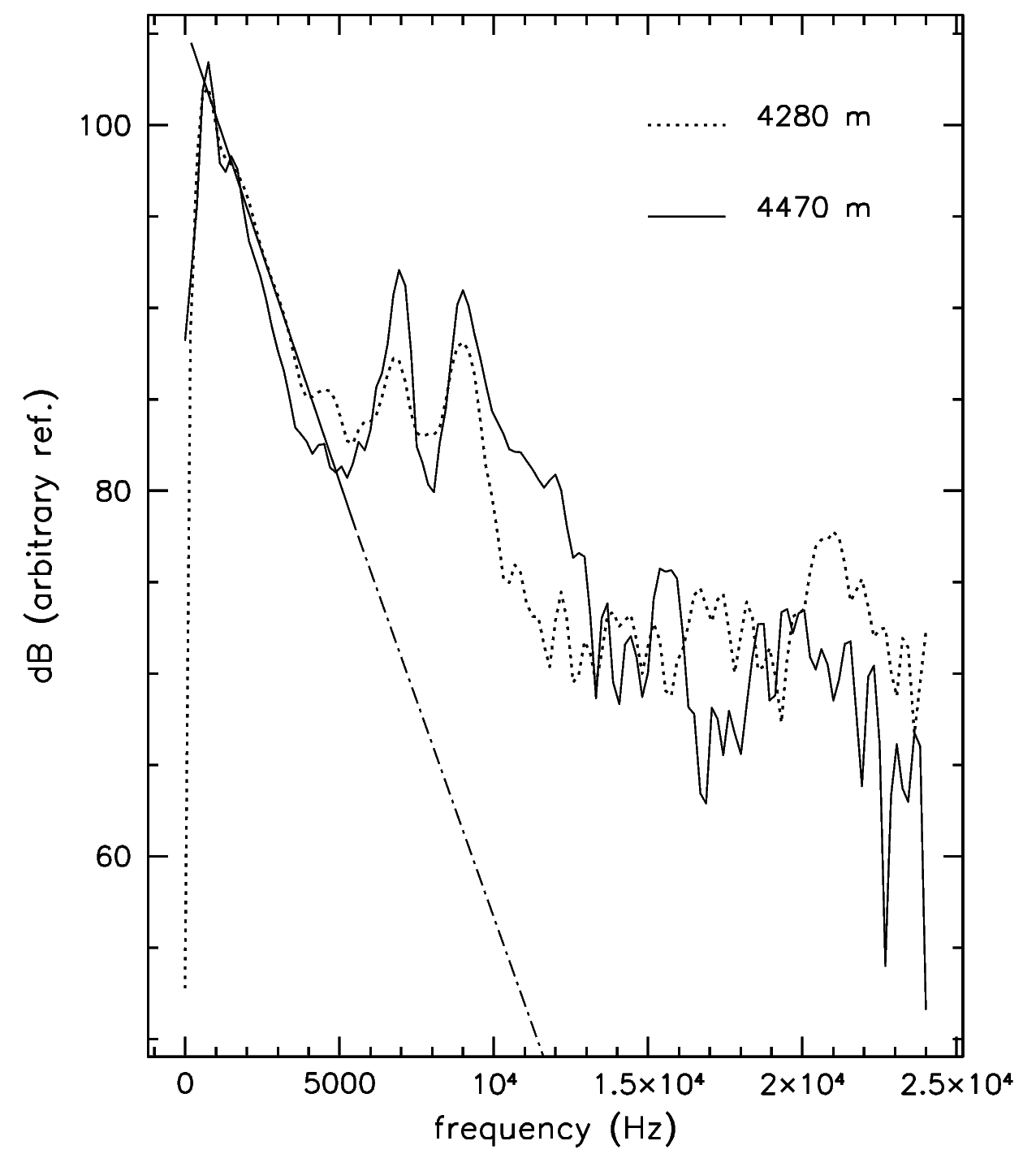}}
\caption{Acoustic power spectrum for the two events, referenced to an
arbitrary baseline power. The data has been corrected for
seawater attenuation over a distance of 4.2 km.
The solid line indicates the 
average slope for Orr and Schoenberg's 1976 data for frequencies below
5 KHz. The dashed extrapolation of this line shows that the higher 
frequency data does not continue this trend. The apparent resonance peaks
in the spectrum at 7 and 9 KHz, and the structure above 10 KHz are
due to hydrophone artifacts, and the slope above 20 KHz is uncorrected
for the falling hydrophone sensitivity.
}
\end{figure}

The broad infall pulse produces the bulk of the lower frequency power,
with a peak of about 600 Hz, slightly higher that that
seen by Orr and Schoenberg (1976). The slope of the decay in the 
spectrum in the region below 5 KHz matches fairly well the average slope
seen in the spectra of Orr and Schoenberg
(solid line), whose spectra cut off at this frequency.
In spite of the artifacts, it is evident that there is significant power
in the shock wave pulse up to the Nyquist limit of the sampling used.
In fact, the flat distribution at higher frequencies shows the impulsive
nature of the shock wave, and indicates that it would be
{ \em unresolved at our sampling interval}, 
if the frequency-dependent water-column attenuation were not present. This flattening may be contrasted with 
results described by Urick (1967), where explosive charges at these depths
show spectra that fall off as power-laws with frequency above 600 Hz.
It appears that implosions on large spheres
may have inherently more high-frequency
spectral content than explosive events of similar energy.

From Figures 5 and 6 it is apparent that
the energetics of the implosion are dominated by the shock wave,
even though the data are uncorrected for water-column
attenuation.  The total acoustic energy for both of our events
is of order 45\% of the available potential
energy of the implosions. This contrasts rather strongly with the
results of Orr and Schoenberg (1976), who found an average of 21\% conversion
to acoustic energy. Since energy goes effectively with the square of the
peak pressure (Urick 1967), the acoustic efficiency will be
underestimated if the shock wave pulse is undersampled.
We suggest that the discrepancy here could be due 
in part to such sampling effects, but may also be related to the fact that 
our implosions took place at depths 20-70\% greater than those of 
Orr and Schoenberg (1976).

\subsection{Mechanical forces}
  
Other than extrapolations of the measured acoustic pressure variation during
the implosion events at the surface, we have two distinct pieces of 
evidence concerning the mechanical forces encountered in the near field.
The first is the yield of the 
steel bolts used to anchor the hardhats into the titanium frame
(and the yield and breaking of the frame itself), and the second is
the shattering of the syntactic float at a distance of
$\sim$ 16 m from the implosion.

\subsubsection{Material yields}

Based on the yields of the various materials used, each hole in the 
titanium frame, and the associated bolts, experienced forces of
order 10 KN during the impulse
which caused the yield. However, there is considerable uncertainty in
converting this to mechanical energy by 
using products of the force and the yield distance, since the 
effective bearing surface area on the parts can only be roughly estimated.
In spite of this uncertainty, a number of estimates by different methods
suggest values as high as $10^6$ Joule for this impulse, 2/3 of the total
potential energy available. 
This exceeds both the estimate from surface measurements and
the expectations from Orr and Schoenberg's results, and supports the
suggestion that significant power in the initial shock is injected into 
very high acoustic frequencies that are absorbed in the near field.

\subsubsection{Shatter of syntactic float}
  
Such high frequency content may explain the apparent shattering of our 
float, although the evidence is less compelling since the history
of the float indicates that it had had many pressure cycles, and
rough handling before our use of it. If we estimate the peak 
overpressure in the shock wave, correcting the surface measurements for
the known transmission loss within our sampling frequency range,
the implied stress is 
\begin{equation}
P_{max} (r)  =   {1 \over {r}}~  {P_{max} (1~\rm{m})} ~10^{ D \alpha (f) / 20 }
\label{stress-eqn}
\end{equation}
where $ P_{max} (1 \rm{m})$ is the peak pressure referenced to 1 m, 
$D$ is the depth in km, and $ \alpha (f)$ is the spectral attenuation 
coefficient in dB $\rm{km}^{-1}$ at frequency $f$ in Hz. The $1/r$ rather
than $1/r^2$ dependence is characteristic of a spherical pressure
wave: since the energy depends on the square of pressure,
and it is the energy fluence which obeys a $1/r^2$ propagation law,
the pressure must fall off as $1/r$.

For the implosion which caused the loss of our float, the peak pressure can be
expected to be at least a factor of three higher than the measured surface
peak pressure (both referenced to 1 m) based on the $\sim 10$ dB loss in 
intensity at the highest frequencies measured, and this is probably
a lower limit. Thus, from equation~\ref{stress-eqn} for a distance of 16 m, 
$ P_{max}   =   4.6 \times 10^6$ Pa (about 660 psi).
However, at 16 m distance, even frequencies of 1 MHz are only attenuated
about 3 dB in the near field water column.(cf. Apel, 1987).
Thus if the indications of the power
spectra in Fig. 7 are correct, this stress estimate is also only a
lower limit. Since even pure cast epoxy has a tensile strength of only
a few thousand psi, the indications are that the pressure stress exceeded
the tensile strength of the float.

Examination of the fractured surfaces of the float showed
granularity and bubbles in the epoxy on scales of order 1 mm,
all of which was apparently part of the construction of the material
(typical sizes of the glass microspheres embedded in the
epoxy matrix are of the order of 0.020 mm or less).
This scale corresponds to a
typical acoustic wavelength at 1 MHz. 
Some of the fractures followed interfaces between
the different molded sections of the float, suggesting that the shock wave
was refracted by the density inhomogeneities within the float. This effect,
combined with the possibility of resonant scattering by inhomogeneities
with wavelengths characteristic of the incoming shock front, suggest
that syntactic foam may be particularly susceptible to such damage.

\subsubsection{Predicted mechanical shock-loading of other components}
  
The damage to the float discussed in the previous experiment raises the
issue of shock-loading of electronics contained within other housings
of the array due to implosions. Two types of housings are now commonly
used for deep ocean arrays for neutrino astronomy: 
optical modules, housed typically in 43 cm diameter glass
instrument housings, with the interior photomultiplier and electronics
coupled mechanically to the housing via a thickness of 
$\sim 1/2$ cm of silicone gel. These modules we assume to be captured
in an external frame of some kind, in our case acrylic.
The second type of housing typicall used is an electronics control 
module (or{\em string controller}. For our purposes here we take the
nominal string controller to be a
20 cm diameter Al tube of approximately 2.5 m length,
with electronics contained in an internal rack-mounting system, with
mechanical isolation bushings between the rigid rack and the tube wall.

To make an estimate of the acceleration experienced by components within these
housings, we treat two limiting cases: first, the resonant acceleration
of an entire housing by an acoustic pressure wave with
wavelength of order twice the 
dimension of the housing. The other extreme is the very high-frequency case
where the wavelengths are short enough to be treated by geometric optics.
The first case should obtain for the underpressure wave and the second for
the highest frequency components of the shock wave.

\paragraph{Resonant acceleration:}
 
For the optical modules, the strongest resonant
coupling will occur from components in the pressure wave with time
scales of $\sim 0.5$ ms, which is the same wavelength as the underpressure
infall wave (not surprisingly, since this is caused by a 43 cm sphere).

For an implosion with a maximum underpressure point of $P_U$ (re 1 m) Pascals,
the resonant acceleration of an OM of the same diameter and mass $M$
at a distance $r$ will be 
\begin{equation}
a_{res} ~(P_U ;  r)   = {1 \over r}~ { {\epsilon ~P_U ~A} \over {M } } 
\label{ares-eqn}
\end{equation}
where the factor $\epsilon \leq 1$ is a coupling efficiency factor,
and the PMT module cross section here is $\sim 0.15$ m.
For $\epsilon  =  50$\%, the acceleration for the deeper of our two
implosions is $1200 ~ \rm{m} ~\rm{s}^{-2}$ (120 g) at a distance of 10 m and
a module mass of 25 kg. This factor is already quite large, probably enough
to damage crystal oscillators permanently. If the coupling efficiency
is near unity, accelerations $\geq 50 $g would be experienced out to ranges
of $\sim 50$ m. This could cripple a substantial part of the array if
no vibration isolation is used for sensitive components. 

Given the survival of the transponder in the first implosion
test at a distance of 6 m from the implosion, there is evidence that
electronics can survive in proximity to such an event. The transponder
was somewhat shielded by the adjacent float spheres, and also the
mechanical coupling to these spheres would provide a damping mass to its 
acceleration, but in any case an acceleration of a few hundred g is 
probable. We are in the process of acquiring documentation on the
transponder electronics to evaluate this further.

The string
mooring tension may help to damp the acceleration somewhat; however, the 
amplitudes of this oscillation are only a few mm even for high coupling
efficiencies at 10 m, so the damping will probably not be large, and
may even induce loads in the string riser elements which couple to other 
modules.

For the string controller, the estimates for bulk resonant coupling are
more uncertain, but we can use the power spectrum in Fig. 7 to scale
the peak of the 600 Hz pressure wave to the resonant frequency of an object of $\sim 2.5$ m length (about 300 Hz). The change from 600 to 300 Hz corresponds
to a change of acoustic power of about -8 dB, or a loss of about a factor
of 2.5 in peak pressure. The expected mass of the string controller is
$\sim 250$ kg, and the area $\sim 0.3 \rm{m}^2$ for a pressure wave traveling
along the string. Thus the acceleration from an implosion at 10 m
distance with a coupling efficiency of 50\% would be $\sim 160 ~ \rm{m} ~\rm{s}^{-2}$, (16 g) which is probably survivable for most components.

\paragraph{Shock acceleration:}
  
The coupling of acoustic energy with wavelengths much smaller than
the scales of the instrument housings proceeds by geometric optics.
For normal incidence, the transmission coefficient per interface
(from medium $i$ into medium $j$) is (cf. Rayleigh 1945):
\begin{equation}
T_{i,j}   =   1   -   \left[ { { {\rho_j}/{\rho_i}  -  {c_i}/{c_j} }
\over { {\rho_j}/{\rho_i}  +  {c_i}/{c_j} } } \right ]^2
\label{trans-eqn}
\end{equation}
where $\rho_i , c_i$ are the density and sound velocity in
medium $i$. For coupling to instrument housings, at least two interfaces are
encountered for coupling to internal electronics, so the total
transmission will be (to first order) the product of the individual 
transmissivities. The coupling of pressure will depend on the
square root of the transmission coefficient, again because of the
relationship between energy and pressure. Thus equation (4) may be rewritten
for the approximate acceleration due to acoustic coupling of the shock wave:
\begin{equation}
a_{shock} ~(P_{max} ;  r) \simeq {1 \over r}~ 
{ \sqrt{ \prod_{i < j} T_{i,j} } }~
{ {P_{max} A_{eff}} \over {M } }
\label{ashock-eqn}
\end{equation}
where $A_{eff}$ is the effective cross section of the module accounting for
the integrals over non-normal angles of incidence, and the product is taken 
over all the interface transmission coefficients.

For the optical module, the seawater-acrylic-seawater transition
has a transmission coefficient of $\sim 95\%$ because acrylic is close in
both density and velocity of sound to water.
The seawater-to-glass transition has a
transmission coefficient of $\sim 0.4$. Unfortunately it is
difficult to estimate the velocity of sound in the
silicone gel that couples the PMT to the sphere. However, 
assuming a speed of sound of $500 ~\rm{m} ~\rm{s}^{-1}$ for this gel, 
the transmission coefficient is $\sim 0.15$.
If we use an effective area of $0.1~ \rm{m}^2$ for the module,
the implied acceleration at 10 m, for a peak overpressure of $10^8$ Pa,
is $10^4 ~\rm{m}~ \rm{s}^2$ (1000 g), ignoring attenuation
of the pressure wave in the gel. Thus it appears that at 10 m
distance, an attenuation of 15-20 dB is required  through this gel
to decrease the shock loading to tolerable levels.

\section{Discussion: Implosion Dynamics}
  
The evidence from the data strongly suggests that the implosion
event forms a fully developed jump-shock, which has the property that
pressure is effectively discontinuous across its surface, leading to
the observed impulsive acoustic event with strong high-frequency components.
The requirement for developing such a shock is that the initial impact
of the infalling material be at velocities in excess of the velocity
of sound in the medium. This then provides the momentum pileup in the
shock front which drives up the local pressure within the front
so that the acoustic velocity in this region is significantly higher
than in the pre-shock fluid. Such conditions thus lead to the
inherent steepening of the leading pressure gradient as the acoustic Fourier
components of the shock are trapped in the high pressure region.
In this section we digress briefly to
consider analytically whether these conditions appear to be fulfilled 
in these implosions.

\subsection{Qualitative description}
  
Before developing analytic results for the implosion parameters,
it is useful to describe the sequence qualitatively. To do this, 
first consider an ideal case where there is no glass shell present,
but a pressure-free
cavity of 43 cm diameter is allowed to collapse. As the collapse
begins, the initial pressure drop begins to propagate radially out
as an acoustic wave. Within this radius, the entire water mass is
beginning to accelerate radially inward, and the continuity equation
indicates that, as the infalling surface of the cavity decreases in area
as $1/r^2$ relative to its initial surface, 
the velocity must increase at the same rate. Thus both the
pressure gradient, and the radial gradient in the surface area 
of the spherical cavity contribute to the
acceleration of the infalling water.

As the water mass meets at the center, an acoustic compression wave is formed
from the momentum of the collision, and this begins to propagate radially
out. However, its velocity is retarded because it is moving
upstream through the still infalling
material above it. Thus,
relative to the onset of the underpressure pulse,
one expects it to be delayed by the infall time, plus the acoustic
velocity out of a sphere radius,
plus an additional delay due to the infall retardation. The 
relaxation time of the infalling material will be of order the acoustic
transit time of the sphere, a few hundred microseconds, and after this time
the outgoing compression wave will continue at the acoustic velocity.
However, during transit of the infall region, the delay can be
substantial if the mean velocity is comparable to the acoustic velocity.
If the infall is supersonic, such effects will lead to a substantial delay
of the outgoing shock wave because the modes are effectively trapped within
the supersonic relaxation region until the infall motion is complete.

\subsection{Semi-analytic Description}
  
Here we analyze the infall of the water into a pressure-free spherical
cavity, initially ignoring the glass shell and assuming spherical symmetry,
and incompressibility for the water. 
The approach is denoted a
semi-analytic one since we will depend somewhat on the characteristics
of the measured profile of the infall pressure. The treatment is not
intended to be accurate for velocities that approach the acoustic 
velocity of water.

\subsubsection{Infall Velocity}
  
We approximate the observed
initial pressure drop (the first 0.5 ms in Figure 4)
of the deeper of our two implosions by a linear gradient in time:
\begin{equation}
P (t)    =    - K  t
\label{pgrad-eqn}
\end{equation}
where the constant $K$ can be evaluated from figure 4 to
be $K \simeq  8 \times 10^9$ Pa $\rm{s}^{-1}$ re 1 m.
Actually, it will be more convenient to reference the dynamical
variables to the initial cavity radius $R_0  =  0.21 $ m, so we
use equation~\ref{stress-eqn} to get $K_0  =  K/R_0  =  3.8 \times 10^{10}$
Pa $\rm{s}^{-1}$.

From Bernoulli's Law we can relate the fluid velocity at $R_0$ to this
measured pressure drop:
\begin{equation}
P (t)   =   - {{ \rho u^2 (t) } \over 2 } 
\label{bernoulli-eqn}
\end{equation}
where the pressure has been referenced to ambient pressure.
Combining equation~\ref{pgrad-eqn} and equation~\ref{bernoulli-eqn},
we find the velocity
as measured at $R_0$:
\begin{equation}
u_0 (t )   =    \sqrt{ {2 K_0 } \over \rho }   ~ t^{1/2}  .
\end{equation}
Just outside the infalling surface (again ignoring
the glass shell) the continuity equation 
(constancy of the product of area and velocity) then implies that
\begin{equation}
u_r (t) =  {{ dr } \over {dt}} = { u_0 (t )}{{ R_0^2 }\over{ r^2 }}
\end{equation}
where the variable $r$ now refers to the radius of the infalling surface.
This provides a separable equation which may be integrated in $t$ and $r$:
\begin{equation}
\int_{R_0}^{r} r^2 ~dr= \sqrt{ {2 K_0 } \over {\rho} } \int_{0}^{t} t^{1/2}~dt 
\end{equation}
The result is:
\begin{equation} 
r (t) = \left[ R_0^3 - 2  R_0^2 \sqrt{{2 K_0} \over{\rho}}~t^{3/2} \right]^{1/3}
\end{equation}
and this may be inverted to get the time $T_c$ that the infall reaches the 
sphere center:
\begin{equation}
T_c   =  \left( {{R_0 } \over 2}  \sqrt{{{\rho} \over{2 K_0 }}} \right)^{2/3} .
\end{equation}

Using the parameters for our second implosion, the infall time is
estimated from this equation to be $\sim 540 ~\mu$s, which is very close to 
what is observed in Fig. 4. At this point the acceleration stops, and
the expansion of the compressed region causes the change in sign of the
first derivative of the 
pressure as observed in Fig. 4. However, as discussed above, the
shock wave will not appear until after 
an additional lag due to relaxation effects, and the transit time out of the sphere. 
The actual leading edge of the shock wave appears at about $1200 ~ \mu$s,
implying that the dynamical effects during the production of the shock
wave after the initial infall are significant, since the transit time out
of the sphere radius is only about $120 ~ \mu$s. 
  
The infall radius and
velocity as functions of time are shown in Figure 8 (a) and (b)
for the deeper of the two implosions of Fig. 4.
The radius at which the velocity becomes supersonic
may be determined by setting the left hand side of equation
(10) equal to $c$ and solving for the radius at the time 
the infall reaches the center. For the example above, the supersonic infall
radius extends out to 7 cm, as is shown by Figure 8(c) which plots
the infall velocity against infall radius.

\begin{figure}
\centerline{\includegraphics[width=3.5in]{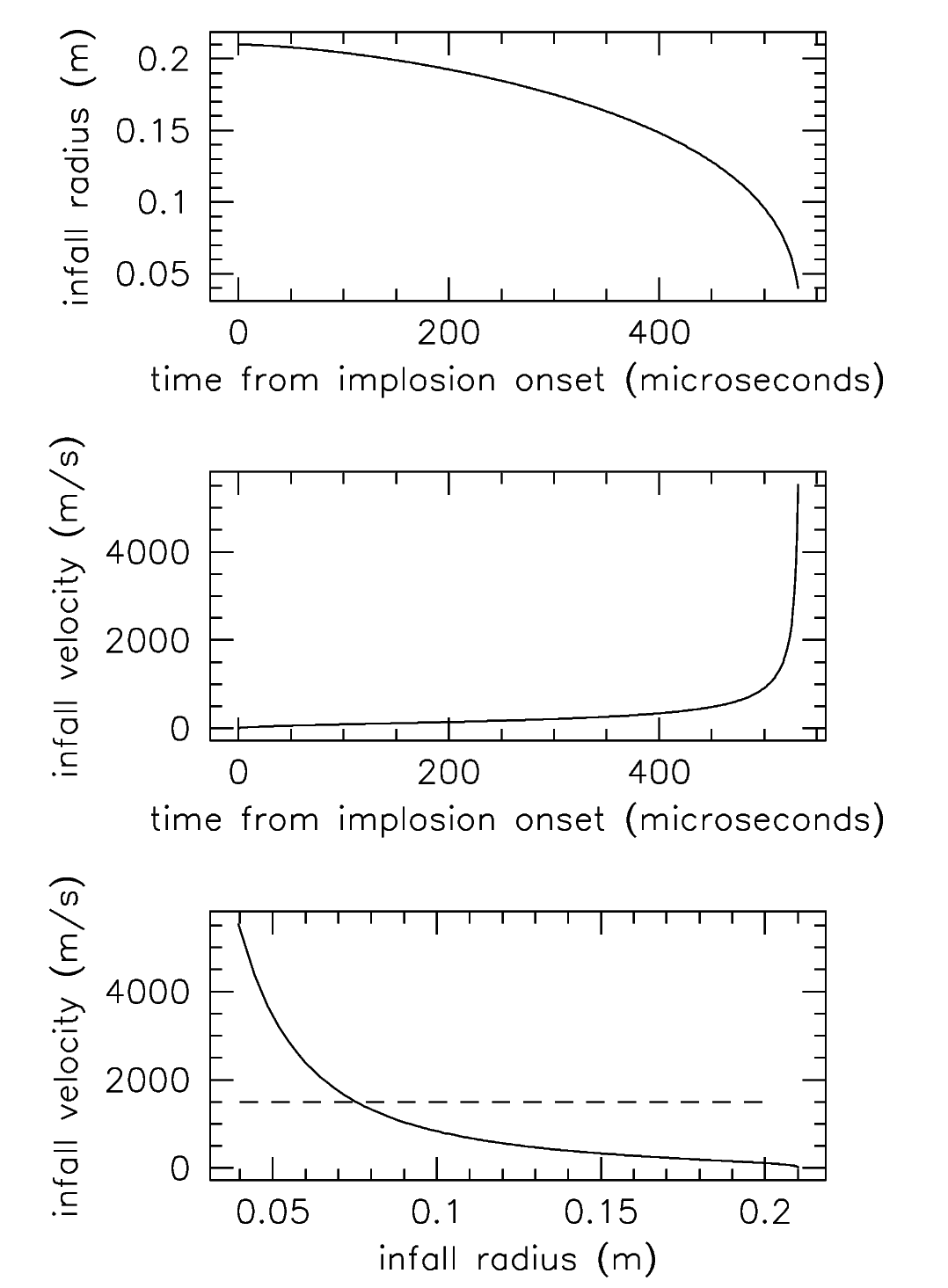}}
\caption{Predicted time dependence of (a) the infall radius, and
(b) the infall velocity, for the deeper of our two implosions.
(c) Infall velocity as a function of infall radius for same implosion.
See text for details of predictive model.
}
\end{figure}

\subsection{Caveats}
  
Obviously this analysis is subject to a number of limitations, which we
mention here. First, we have ignored effects of the compressibility of
water; however, such effects will be of critical importance in the
development of the shock wave. Compressibility plays a role in the
regions of subsonic velocity as well, leading to some rebound effects
in the initial underpressure wave. However, for our purposes, which are
primarily to demonstrate that conditions for shock formation are
present, water may be considered incompressible to first order.

Two other effects which are ignored in this initial treatment are 
the glass shell, and the internal pressure of the sphere. We discuss
both in some detail in the remainder of this section.

\subsubsection{Effects of glass shell}
  
The velocity of shear waves in pyrex glass 
at atmospheric pressure is $\sim 3300  ~\rm{m} ~\rm{s}^{-1}$
and therefore a 
time scale of $\sim 0.2$ ms is required for fractures to propagate
from one side of the sphere to the other. 
Thus the glass initially provides some resistance to the infall while the 
fracturing process continues. This effect may be the cause of the
initially lower gradient in the pressure function seen in Fig. 4.
There is a distinct change in slope at a time of $\sim 160~ \mu$s after
the infall onset, of the right order of magnitude for the 
initial stages of the glass fracturing event.

Once the fracturing has crossed the sphere, no pressure
support can be maintained, and the implosion should proceed in a manner
similar to that described above, with the glass being driven out
in front of the infalling water surface. However,
the higher density of the glass as compared to water will also slow down
the infall somewhat, since the velocity depends on $\rho^{-1/2}$.
We have made estimates of the density of glass agglomerations recovered from
the remains of hardhats that survived sphere implosions similar to ours,
and these fall in the range $1.2-1.7 ~\rm{gm}~\rm{cm}^{-3}$. Thus the glass 
might be expected to impede the infall rate by $\sim 30-40\%$, if all
of the infall volume were an admixture of glass and water. However,
the glass only comprises 1/5 of the sphere volume, and thus the
square root of the average density only increases by $\sim 10\%$.

Another likely result of the fracturing event is to introduce 
asymmetry in the infall geometry, but it was found by Orr and Schoenberg
(1976) that the spheres imploded in pressure vessels yielded glass which
was completely fractured to sand-sized particles.
Glass recovered from
deep-ocean implosions shows a particle size spectrum ranging from sand
down to micron sizes. The absence of
large pieces of glass suggests that enough symmetry is maintained during
the implosion so that nearly complete fracturing can still take place.

However, the volume of the glass involved is sufficient to produce
a sphere of radius $\sim 13$ cm at the center of the implosion if it retained
its pre-fracture density, which is not the case. Using the average 
of the densities
that we have reported above, the volume of the fractured glass corresponds
to a sphere of order 15 cm radius, 3/4 that of the entire pre-implosion radius.
Thus the presence of the glass is clearly important to the dynamics of 
the shock wave formation, and further complicates interpretation of
the pressure signature, since it occupies the volume in
which the shock wave is generated.

We note in passing that the potential energy of
the event is a small fraction of what would be required to
fuse this volume of glass, even if the 
conversion to thermal energy were 100\% efficient.

\subsubsection{Effects of internal pressure}
  
The gas internal to the sphere was neglected above in calculating the
infall velocities.
Here we assess its effects on the implosion dynamics.

The critical radius $r_{C}$  at which internal pressure effects 
can be expected to become important is of order
\begin{equation}
{r_{C} } = { R_0 } \left( {{ P_{I} } \over{P_A}}\right)^{1/4}
\end{equation}
where we have assumed the ideal gas law with adiabatic index $\gamma  =  4/3$,
the initial internal pressure is $P_I$, and $P_A$ is the
ambient pressure ($P_A \simeq D \times 10^7$ Pa at $D$ km depth).

As an example, if the sphere had an initial internal pressure
$P_I = 1$ atm, the compressed internal gas would reach 20\% of the
ambient pressure for 4.5 km depth at a radius of $\sim 7$ cm,
and would equal the ambient pressure at $\sim 4.5$ cm. Thus 
an internal pressure of 1 atm might lead to some moderation of the
infall velocity near the supersonic radius, possibly decreasing it
somewhat. Initial internal pressures in our tests were 0.22 atm
and we conclude that they had probably had little effect on the dynamics
of the shock formation, as compared to effects due to the matrix
of fractured glass the occupies the central volume as the shock wave begins 
to form. 
It is worth noting that Orr and Schoenberg used two different internal
pressures in the spheres they studied, 0.7 atm and 0.07 atm, and they
could see no difference in the behavior of the acoustic signature of the
two different types, a result that supports our conclusion above.

\section{Conclusions}
  
A rigid titanium frame was successfully used to protect 
tensile and electro-optical cables in a riser bundle similar to that 
used in a typical deep ocean array, from damage due to the
energetic implosion of a 43 cm Benthos Corp. glass instrument housing.
The frame itself was damaged in its weakest member, and the explosive
forces present were strong enough to cause yield of portions of the
frame, but the protective function of the frame was still intact,
and its strength as part of the tensile riser of the mooring used
was also retained. It is recommended that the crossmember of the frame
be strengthened since it is preferable to have the frame remain intact
to avoid having the sharp surfaces of the broken frame near the
kevlar riser following an implosion event.

We have indirect evidence that similar glass housings within a distance
of $\sim 6$ m of an imploding sphere did not undergo sympathetic 
implosions, and direct evidence that a sphere within 24 m of an imploding
sphere remained intact.
Since some of the distances between spheres within deep ocean arrays
could be less than 6 m, further tests should be done to determine the
safe distance at which a second sphere can survive.

A syntactic float within 16 m of an implosion was apparently shattered
by the event, possible due to much higher near-field pressure gradients
than are measured at the surface. The surface acoustic profiles do show
power up to the Nyquist frequency due to the shock-wave pulse, indicating that
higher frequencies were also present. A simple qualitative analysis of
the development of the shock wave also indicates that shock conditions,
requiring supersonic collision of the infalling material, are strongly
satisfied in such an implosion event. This analysis also supports
the hypothesis that the near field pressure peak may be dominated
by very high frequencies which are absorbed before reaching the surface.
Thus it is recommended that the top float for each string be separated to
a distance of $\gg 16$ m above the last glass sphere. The use of glass
floats in the array should be considered as well, but the problem of
sympathetic implosions should then be investigated in more detail.
Mechanical isolation of the most shock-sensitive components in the
optical modules appears to be necessary if they are to survive the implosion
shock within a distance of a few tens of meters or more.

Because of the probable presence of very high frequencies in the near
field around an implosion event, it may be worthwhile to develop
high frequency attenuation masks around hydrophones in such an array,
for frequencies above the Nyquist sampling limit of the recording 
system. The maximum overpressure that is possible in the near field
event could correspond to an increase in depth of up
to 8 km momentarily; thus the peak pressure may exceed the rating of 
the hydrophones, and an attenuator may help reduce this.

\acknowledgements
  
We wish to thank Sam Raymond of Benthos Corporation for his
invaluable assistance in preparing the implosion spheres for this
experiment. We also thank Capt. Jack Ross, skipper of the 
vessel {\em Capt. Jack}, of Keauhou Bay, Hawaii Island, for his help
in staging the deployment operations, his amazing stamina,
and continual good cheer.

\subsection*{References}

Apel, J. R., 1987. Principles of Ocean Physics. (London: Academic Press).
\newline
\newline
O'Connor, D. J., 1990. Unpublished Dissertation, University of Hawaii.
\newline
\newline
Orr, M., and Schoenberg, M., 1976. Journ. Acous. Soc. Am. 59, 1155-1159.
\newline
\newline
Raymond, S. O., 1975. IEEE Ocean `75, 537-544.
\newline
\newline
Lord Rayleigh, 1945. The Theory of Sound, vol II, (New York: Dover).
\newline
\newline
Urick, R. J., 1967. Principles of Underwater Sound for Engineers.
(New York: McGraw-Hill).

\end{document}